\documentclass[sigconf]{acmart}

\usepackage{booktabs}
\usepackage{tabularx}
\usepackage{array}
\usepackage{multirow}

\usepackage{tikz}
\usetikzlibrary{arrows.meta,positioning,fit,calc}

\setcopyright{none}
\renewcommand\footnotetextcopyrightpermission[1]{}

\acmConference[]{}{}{}
\acmYear{}
\acmDOI{}

\begin{document}

\title{Cost--Utility Alignment in LLM Agent Trajectories: Profiling, Attribution, Diagnosis, Adaptation, and Evaluation}

\author{Dan Liu}
\email{202621081050@mail.bnu.edu.cn}
\affiliation{%
  \institution{School of Artificial Intelligence, Beijing Normal University}
  \city{Beijing}
  \country{China}
}

\author{Jian Li}
\email{jli@mail.bnu.edu.cn}
\affiliation{%
  \institution{School of Artificial Intelligence, Beijing Normal University}
  \city{Beijing}
  \country{China}
}

\begin{abstract}
LLM agents execute tasks through multi-step trajectories that accumulate cost in tokens, latency, monetary fees, and environmental risk while producing utility only at the aggregate task level. Prior surveys address inference optimization, agent capabilities, or evaluation in isolation, leaving practitioners without principled tools to determine whether a trajectory's resource expenditure is justified by its task contribution. We address this gap by developing a trajectory-centric cost--utility alignment framework that treats resource consumption and task contribution as dual ledgers over the same execution, organized around five analytical stages: cost profiling, utility attribution, misalignment diagnosis, targeted adaptation, and evaluation. Utility attribution is central to this structure: rather than relying on aggregate outcomes, it organizes contribution methods by evidential strength, from process proxies and information dependency to counterfactual replay, supplying the causal evidence that grounds diagnosis and guides adaptation. Using this framework, we analyze recent agent systems, attribution methods, and evaluation protocols covering efficiency, reliability, and economic value, as well as five forms of misalignment spanning cognitive and context use, external interaction, recovery-loop control, resource--capability allocation, and multi-agent coordination, together with their targeted adaptations. The result is a closed analytical loop connecting the cost side of agent execution to its utility side, providing a structured basis for resource-aware agent design and deployment.
\end{abstract}



\maketitle
\pagestyle{plain}

\section{Introduction}

LLM agents execute tasks through stateful trajectories, typed event sequences that accumulate model calls, tool invocations, environmental actions, and observations across turns toward a task objective~\cite{arxiv2308_11432}. The trajectory, not the individual model call, is the appropriate unit of efficiency analysis, with costs compounding as context is repeatedly processed, tools are called, and failures are recovered, while utility emerges from the aggregate task outcome. Deployed across web, software-engineering, desktop, and command-line environments~\cite{arxiv2307_13854,arxiv2310_06770,arxiv2404_07972,arxiv2601_11868}, these agents make cost and utility trajectory-level properties.

LLM agent trajectories create a deployment tension in which additional reasoning, retrieval, verification, and collaboration raise task performance alongside tokens, latency, tool fees, human supervision, and exposure to external side effects. Token accounting alone captures only part of this burden. Repeated context dominates many agent bills, long contexts and serial calls create system costs beyond nominal tokens, and inexpensive tool actions can still produce irreversible environmental loss~\cite{arxiv2604_22750,arxiv2506_04301,ruan2024toolemu}. Difficult tasks require extensive exploration, strict verification, or recovery from genuine faults, so interpreting cost against the utility of the same execution is essential.

Existing work addresses inference optimization, agent capabilities, task evaluation, safety, or memory mechanisms in isolation, leaving resource consumption and task contribution unconnected~\cite{arxiv2308_11432,arxiv2404_14294,arxiv2503_16416}. We organize both sides under a cost--utility alignment framework, introducing utility attribution as the analytical counterpart to cost profiling and connecting the two through misalignment diagnosis, targeted adaptation, and evaluation into a closed analytical loop.

(1) We propose a cost--utility alignment framework built on a typed agent trajectory and a five-stage analytical loop covering cost profiling, utility attribution, misalignment diagnosis, targeted adaptation, and evaluation. The framework treats cost and contribution as dual ledgers over the same execution and defines alignment as the absence of a feasible alternative achieving comparable utility at lower cost under matched constraints.

(2) We position utility attribution as the analytical counterpart to cost profiling, organizing contribution methods by evidential strength from process proxies and information dependency to counterfactual replay, and connecting attribution evidence to misalignment diagnosis, targeted adaptation, and cost--utility evaluation.

This paper develops a unified analytical framework for resource-aware LLM agents and uses it to organize recent advances in cost--utility alignment across model inference, tools, memory, environmental interaction, and inter-agent communication. Section~2 defines the framework, Sections~3--6 cover the five analytical stages, and Sections~7--8 identify challenges and conclude.

\begin{figure*}[t!]
  \centering
  % Editable source: figures/figure1-organization-architecture.drawio.svg
  % High-resolution raster export is used here for reliable pdflatex compilation.
  \includegraphics[width=0.90\textwidth,keepaspectratio]{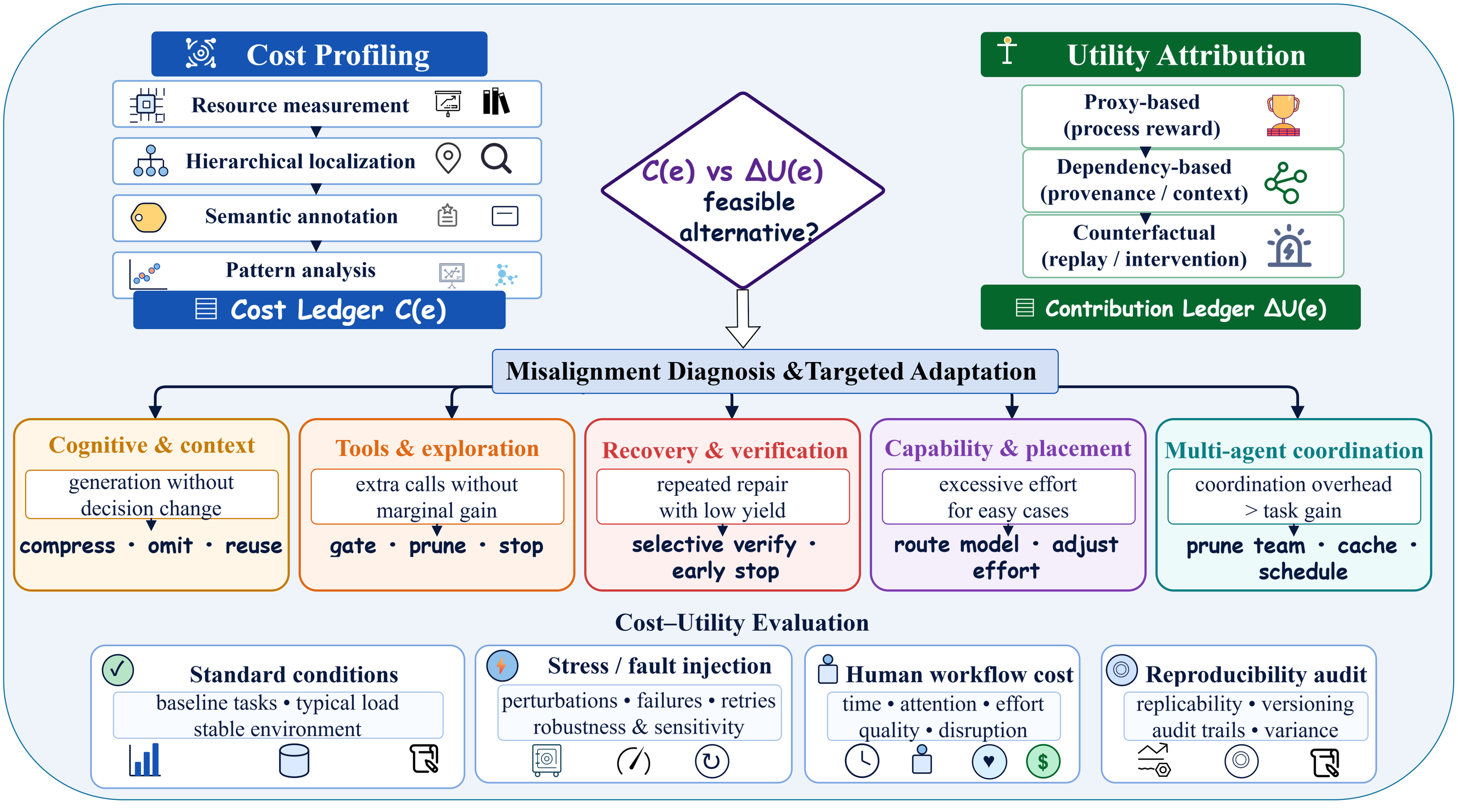}
  \caption{The cost--utility alignment framework.}
  \Description{The five-stage workflow connects cost profiling, utility attribution, misalignment diagnosis, targeted adaptation, and cost--utility evaluation, with evaluation results feeding the next analysis cycle.}
  \label{fig:cost-utility-workflow}
\end{figure*}

\section{The Cost--Utility Alignment Framework}

We propose a five-stage cost--utility alignment framework connecting cost profiling, utility attribution, misalignment diagnosis, targeted adaptation, and evaluation into a closed analytical loop. The framework operates on \textbf{agent trajectories} $\tau=(e_1,\ldots,e_n)$, typed hierarchical event sequences linked by execution order and data flow. Each event $e_i$ carries a \textbf{cost vector} $C(e_i)=(r_i,m_i,\rho_i)$ recording resource consumption, monetary cost, and risk as independent dimensions, with trajectory cost $C(\tau)=\sum_{i=1}^n C(e_i)$. At the run level, a \textbf{utility signal} $U(\tau)$ summarizes goal completion, partial progress, constraint satisfaction, and output quality. \textbf{Local contribution} $\Delta U(e_i)=U(\tau)-U(\tau[e_i{\to}\varnothing])$ connects the two by measuring the utility change when $e_i$ is replaced by a counterfactual baseline. A \textbf{misalignment condition} holds when a feasible $\tau'$ satisfies $U(\tau')\geq U(\tau)$ at $C(\tau')<C(\tau)$, or $U(\tau')>U(\tau)$ at $C(\tau')\leq C(\tau)$, within \textbf{budget constraint} $C(\tau)\leq B$.

\begin{figure*}[t]
  \centering
  \resizebox{\textwidth}{!}{\begin{tikzpicture}[
  font=\sffamily,
  chapter/.style={
    draw=black!65, rounded corners=1.2mm, fill=yellow!7,
    line width=0.35pt, text width=25mm, minimum height=8mm,
    align=center, inner xsep=1.6mm, inner ysep=0.9mm,
    font=\fontsize{8.0}{8.8}\selectfont\bfseries
  },
  topic/.style={
    draw=black!65, rounded corners=1.2mm, fill=black!6,
    line width=0.35pt, text width=38mm, minimum height=8mm,
    align=center, inner xsep=1.4mm, inner ysep=0.8mm,
    font=\fontsize{7.6}{8.4}\selectfont
  },
  papers/.style={
    draw=black!55, rounded corners=1.2mm, fill=green!7,
    line width=0.35pt, text width=100mm, minimum height=8mm,
    align=center, inner xsep=1.4mm, inner ysep=0.75mm,
    font=\fontsize{7.2}{8.0}\selectfont
  },
  rootlabel/.style={
    draw=black!70, rounded corners=1.2mm, fill=yellow!5,
    line width=0.4pt, minimum width=7mm, minimum height=38mm,
    align=center, inner sep=1.4mm, font=\fontsize{8.0}{8.8}\selectfont\bfseries
  },
  branch/.style={draw=black!70, line width=0.4pt},
  node distance=0.8mm and 3mm
]

% Cost profiling
\node[topic] (s31) {Multi-Dimensional Resource Profiling};
\node[papers, right=3mm of s31] (p31) {
  \textbf{AI Agent Token Consumption}~\cite{arxiv2604_22750},
  \textbf{ProMCP}~\cite{anjum2026promcp},
  \textbf{Dynamic Reasoning Cost}~\cite{arxiv2506_04301},
  \textbf{Budget-Aware Tool-Use}~\cite{arxiv2511_17006},
  \textbf{ToolEmu}~\cite{ruan2024toolemu}
};
\node[topic, below=of s31] (s32) {Hierarchical Cost Localization};
\node[papers, right=3mm of s32] (p32) {
  \textbf{Tokenomics}~\cite{arxiv2601_14470},
  \textbf{Terminal-Bench}~\cite{arxiv2601_11868},
  \textbf{ClawTrace}~\cite{arxiv2604_23853}
};
\node[topic, below=of s32] (s33) {Semantic Action Annotation};
\node[papers, right=3mm of s33] (p33) {
  \textbf{SWE-agent}~\cite{arxiv2405_15793},
  \textbf{CostBench}~\cite{liu2026costbench},
  \textbf{AppWorld}~\cite{arxiv2407_18901},
  \textbf{Revisable by Design}~\cite{arxiv2604_23283}
};
\node[topic, below=of s33] (s34) {Workload Characterization and Cost Pattern Analysis};
\node[papers, right=3mm of s34] (p34) {
  \textbf{TraceLab}~\cite{arxiv2606_30560},
  \textbf{AI Agent Token Consumption}~\cite{arxiv2604_22750},
  \textbf{Tokenomics}~\cite{arxiv2601_14470},
  \textbf{Terminal-Bench}~\cite{arxiv2601_11868}
};
\node[fit=(s31)(s34), inner sep=0pt] (g3) {};
\node[chapter, left=3mm of g3] (c3) {Cost Profiling};

% Utility attribution
\node[topic, below=2.2mm of s34] (s41) {Proxy-Based Attribution};
\node[papers, right=3mm of s41] (p41) {
  \textbf{IPR}~\cite{xiong2024watch},
  \textbf{AgentPRM/InversePRM}~\cite{arxiv2502_10325},
  \textbf{iStar}~\cite{arxiv2509_19199},
  \textbf{PABU}~\cite{arxiv2602_09138}
};
\node[topic, below=of s41] (s42) {Information Dependency and Evidence Contribution};
\node[papers, right=3mm of s42] (p42) {
  \textbf{TRACER}~\cite{arxiv2605_09934},
  \textbf{NeuroTaint}~\cite{arxiv2604_23374},
  \textbf{ContextCite}~\cite{arxiv2409_00729}
};
\node[topic, below=of s42] (s43) {Intervention-Based Counterfactual Attribution};
\node[papers, right=3mm of s43] (p43) {
  \textbf{CAR}~\cite{arxiv2606_08275},
  \textbf{ErrorProbe}~\cite{arxiv2606_01365},
  \textbf{CausalFlow}~\cite{arxiv2605_25338},
  \textbf{C3}~\cite{arxiv2603_06859}
};
\node[fit=(s41)(s43), inner sep=0pt] (g4) {};
\node[chapter, left=3mm of g4] (c4) {Utility Attribution};

% Misalignment diagnosis and targeted adaptation
\node[topic, below=2.2mm of s43] (s51) {Cognitive and Context Misalignment};
\node[papers, right=3mm of s51] (p51) {
  \textbf{DEPO}~\cite{chen2026depo},
  \textbf{Agent-Omit}~\cite{arxiv2602_04284},
  \textbf{AgentDiet}~\cite{arxiv2509_23586},
  \textbf{Beyond Compaction}~\cite{arxiv2606_11213},
  \textbf{PRISM}~\cite{arxiv2605_12260},
  \textbf{PEEK}~\cite{arxiv2605_19932}
};
\node[topic, below=of s51] (s52) {External Interaction Misalignment};
\node[papers, right=3mm of s52] (p52) {
  \textbf{Tools-and-Planning Benchmark}~\cite{arxiv2601_02663},
  \textbf{Calibrate-Then-Act}~\cite{arxiv2602_16699},
  \textbf{INTENT}~\cite{liu2026intent},
  \textbf{ToolTree}~\cite{arxiv2603_12740},
  \textbf{W\&D}~\cite{arxiv2602_07359},
  \textbf{Budget-Aware Tool-Use}~\cite{arxiv2511_17006},
  \textbf{WebRollback}~\cite{arxiv2504_11788}
};
\node[topic, below=of s52] (s53) {Recovery-Loop and Verification-Control Misalignment};
\node[papers, right=3mm of s53] (p53) {
  \textbf{ToolMaze}~\cite{arxiv2606_05806},
  \textbf{WAREX}~\cite{kara2025warex},
  \textbf{Self-Repair Study}~\cite{arxiv2306_09896},
  \textbf{Structured Feedback}~\cite{arxiv2607_14167},
  \textbf{GA-Rollback}~\cite{arxiv2503_02519},
  \textbf{EET}~\cite{arxiv2601_05777},
  \textbf{VRR-Stop}~\cite{arxiv2607_17641}
};
\node[topic, below=of s53] (s54) {Resource--Capability Misalignment};
\node[papers, right=3mm of s54] (p54) {
  \textbf{Hybrid LLM}~\cite{arxiv2404_14618},
  \textbf{RouteLLM}~\cite{arxiv2406_18665},
  \textbf{BEST-Route}~\cite{arxiv2506_22716},
  \textbf{Ares}~\cite{arxiv2603_07915},
  \textbf{Inference-Time Distillation}~\cite{arxiv2512_02543},
  \textbf{Hera}~\cite{arxiv2605_24598},
  \textbf{UniScale}~\cite{arxiv2605_30898}
};
\node[topic, below=of s54] (s55) {Multi-Agent Coordination Misalignment};
\node[papers, right=3mm of s55] (p55) {
  \textbf{MAST}~\cite{arxiv2503_13657},
  \textbf{MaAS}~\cite{arxiv2502_04180},
  \textbf{Nash-CredMAS}~\cite{fan2026nashcredmas},
  \textbf{Shapley-Coop}~\cite{arxiv2506_07388},
  \textbf{AgentPrune}~\cite{arxiv2410_02506},
  \textbf{Evolving Orchestration}~\cite{arxiv2505_19591},
  \textbf{SupervisorAgent}~\cite{lin2026stopwastingtokensefficient},
  \textbf{KVCOMM}~\cite{arxiv2510_12872},
  \textbf{Continuum}~\cite{arxiv2511_02230}
};
\node[fit=(s51)(s55), inner sep=0pt] (g5) {};
\node[chapter, left=3mm of g5] (c5) {Cost--Utility\\Misalignment\\\& Targeted Adaptation};

% Cost--utility evaluation
\node[topic, below=2.2mm of s55] (s61) {Cost-Efficiency Metrics and Quality--Cost Trade-offs};
\node[papers, right=3mm of s61] (p61) {
  \textbf{ScienceAgentBench}~\cite{arxiv2410_05080},
  \textbf{FDABench}~\cite{arxiv2509_02473},
  \textbf{SWE-Effi}~\cite{fan2025sweeffi},
  \textbf{AgencyBench}~\cite{li2026agencybench},
  \textbf{Agentless}~\cite{arxiv2407_01489},
  \textbf{Tools-and-Planning Benchmark}~\cite{arxiv2601_02663},
  \textbf{TPS-Bench}~\cite{arxiv2511_01527},
  \textbf{CostBench}~\cite{liu2026costbench},
  \textbf{CUJBench}~\cite{arxiv2604_23455},
  \textbf{Terminal-Bench}~\cite{arxiv2601_11868}
};
\node[topic, below=of s61] (s62) {Reliability and Risk-Adjusted Cost Evaluation};
\node[papers, right=3mm of s62] (p62) {
  \textbf{$\tau$-bench}~\cite{arxiv2406_12045},
  \textbf{ReliabilityBench}~\cite{arxiv2601_06112},
  \textbf{APEX-Agents}~\cite{arxiv2601_14242},
  \textbf{WirelessBench}~\cite{arxiv2603_21251},
  \textbf{CCPO}~\cite{si2026ccpo}
};
\node[topic, below=of s62] (s63) {Human and Economic Value};
\node[papers, right=3mm of s63] (p63) {
  \textbf{HCAST}~\cite{arxiv2503_17354},
  \textbf{PaperBench}~\cite{arxiv2504_01848},
  \textbf{GDPval}~\cite{arxiv2510_04374},
  \textbf{SWE-Lancer}~\cite{arxiv2502_12115}
};
\node[topic, below=of s63] (s64) {Benchmark Infrastructure and Evaluation Protocols};
\node[papers, right=3mm of s64] (p64) {
  \textbf{AgentBoard}~\cite{arxiv2401_13178},
  \textbf{WebArena}~\cite{arxiv2307_13854},
  \textbf{VisualWebArena}~\cite{arxiv2401_13649},
  \textbf{WorkArena}~\cite{arxiv2403_07718},
  \textbf{OSWorld}~\cite{arxiv2404_07972},
  \textbf{SWE-bench}~\cite{arxiv2310_06770},
  \textbf{SpreadsheetBench}~\cite{arxiv2406_14991},
  \textbf{Terminal-Bench}~\cite{arxiv2601_11868},
  \textbf{BenchGuard}~\cite{arxiv2604_24955},
  \textbf{PaperBench}~\cite{arxiv2504_01848}
};
\node[fit=(s61)(s64), inner sep=0pt] (g6) {};
\node[chapter, left=3mm of g6] (c6) {Cost--Utility Evaluation};

% Root and orthogonal branches
\node[fit=(c3)(c6), inner sep=0pt] (allchapters) {};
% Rotate only the text box, rather than the TikZ node itself.  This keeps
% root.east on the physical right edge, so the main branch cannot cross the node.
\node[rootlabel, left=3mm of allchapters] (root)
  {\rotatebox{90}{Cost--Utility Alignment}};

\foreach \source/\target in {
  c3/s31,c3/s32,c3/s33,c3/s34,
  c4/s41,c4/s42,c4/s43,
  c5/s51,c5/s52,c5/s53,c5/s54,c5/s55,
  c6/s61,c6/s62,c6/s63,c6/s64}
  \draw[branch] (\source.east) -- ++(1.5mm,0) |- (\target.west);

\foreach \source/\target in {
  s31/p31,s32/p32,s33/p33,s34/p34,
  s41/p41,s42/p42,s43/p43,
  s51/p51,s52/p52,s53/p53,s54/p54,s55/p55,
  s61/p61,s62/p62,s63/p63,s64/p64}
  \draw[branch] (\source.east) -- (\target.west);

\foreach \target in {c3,c4,c5,c6}
  \draw[branch] (root.east) -- ++(1.5mm,0) |- (\target.west);

\end{tikzpicture}}
  \caption{Taxonomy of methods for cost--utility alignment in LLM agents.}
  \Description{A hierarchical taxonomy organizing representative methods into cost profiling, utility attribution, cost--utility misalignment and targeted adaptation, and cost--utility evaluation.}
  \label{fig:alignment-framework}
\end{figure*}
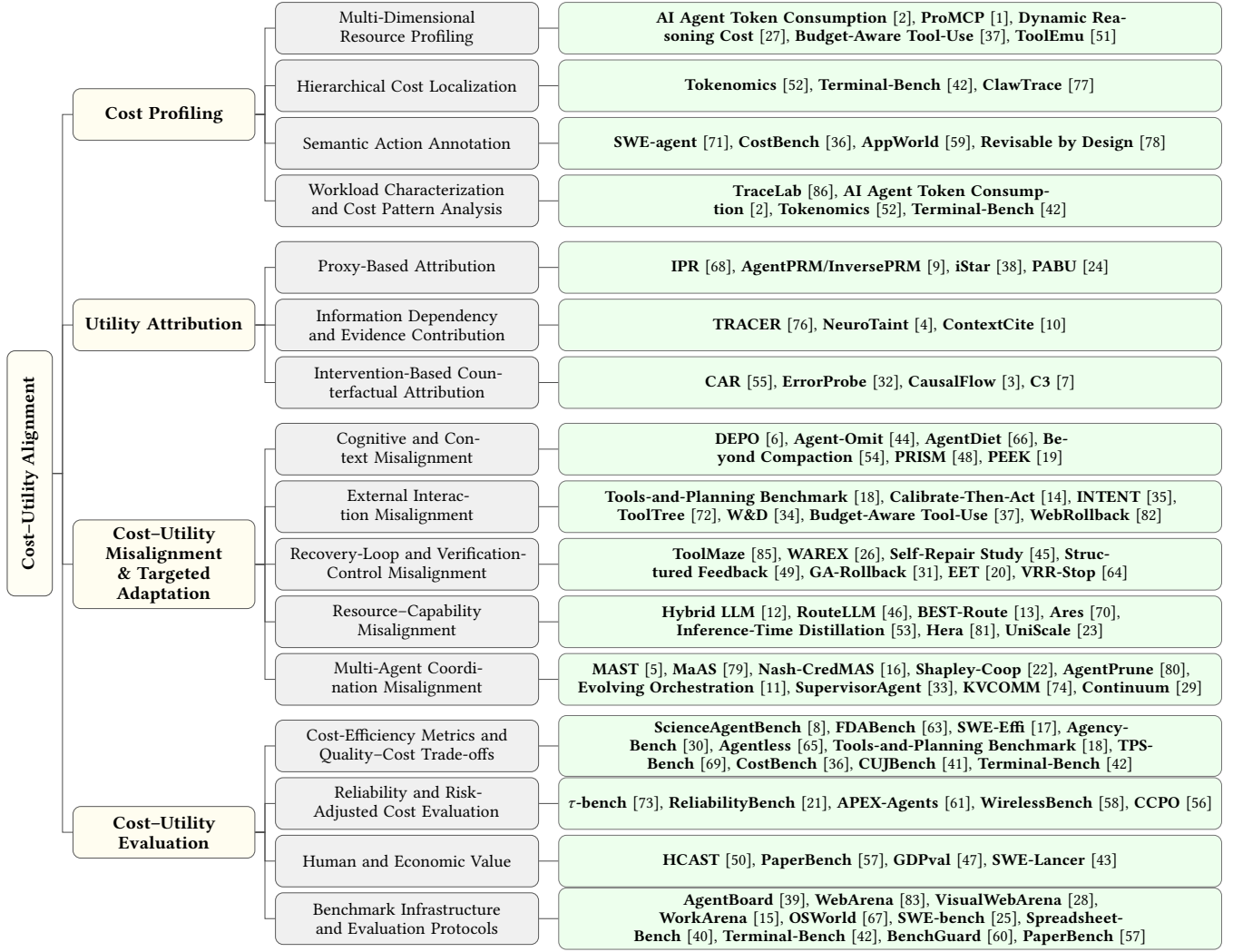

The framework connects two parallel analytical views over the same trajectory and closes them through adaptation and evaluation. Cost profiling builds the resource side, localizing consumption to events and actors, adding semantic labels, and characterizing workload distributions, while utility attribution builds the contribution side, identifying which states, actions, and members are causally responsible for task outcomes. With both ledgers in hand, misalignment diagnosis tests whether high-cost behavior has sufficient utility support relative to a feasible baseline, and targeted adaptation responds by changing the responsible mechanism. Cost--utility evaluation closes the loop by executing the revised system under matched conditions and returning evidence to the next cycle. Figures~\ref{fig:cost-utility-workflow}--\ref{fig:alignment-framework} and Table~\ref{tab:method-taxonomy} illustrate the closed loop and taxonomy; Appendix~A provides a field-level TraceCard schema and a contrasting event-pair instance demonstrating how attribution evidence drives misalignment diagnosis.

\begin{table*}[t]
  \caption{Representative papers in the five-stage cost--utility alignment taxonomy. Each row provides one non-exhaustive example per subsection; an em dash denotes a dimension outside the paper's primary focus.}
  \Description{A centered three-line taxonomy table containing one representative paper for each of the sixteen substantive subsections. Columns compare cost type, utility signal, attribution granularity, adaptation method, and evaluation metrics. The selection is illustrative rather than exhaustive.}
  \label{tab:method-taxonomy}
  \centering
  \scriptsize
  \setlength{\tabcolsep}{1.25pt}
  \renewcommand{\arraystretch}{1.05}
  \renewcommand{\tabularxcolumn}[1]{m{#1}}
  \begin{tabularx}{\textwidth}{>{\centering\arraybackslash}m{0.078\textwidth}>{\centering\arraybackslash}m{0.155\textwidth}>{\centering\arraybackslash}m{0.09\textwidth}>{\centering\arraybackslash}m{0.125\textwidth}>{\centering\arraybackslash}m{0.13\textwidth}>{\centering\arraybackslash}m{0.115\textwidth}>{\centering\arraybackslash}m{0.145\textwidth}>{\centering\arraybackslash}X}
    \toprule
    \textbf{Stage} & \textbf{Subsection category} & \textbf{Representative paper} & \textbf{Cost type} & \textbf{Utility signal} & \textbf{Attribution granularity} & \textbf{Adaptation method} & \textbf{Evaluation metrics} \\
    \midrule
    \multirow{4}{=}{\centering\textbf{Stage 1}\par Cost Profiling}
      & Multi-Dimensional Resource Profiling
      & ProMCP~\cite{anjum2026promcp}
      & Tokens, tool-protocol overhead, latency, and monetary cost
      & \textemdash
      & Protocol stage and tool call
      & \textemdash
      & Token, latency, and monetary overhead \\
    \addlinespace[1pt]
      & Hierarchical Cost Localization
      & ClawTrace~\cite{arxiv2604_23853}
      & Model, tool, and sub-agent call cost and latency
      & \textemdash
      & Task, run, span, and actor
      & \textemdash
      & Per-call cost and latency-path localization \\
    \addlinespace[1pt]
      & Semantic Action Annotation
      & CostBench~\cite{liu2026costbench}
      & Model-token and environment-action cost
      & Task completion under costs and preferences
      & Action and complete trajectory
      & Replanning after cost or environment changes
      & Success, action-path cost, and total cost \\
    \addlinespace[1pt]
      & Workload Characterization and Cost Pattern Analysis
      & TraceLab~\cite{arxiv2606_30560}
      & Context tokens, cached prefixes, and estimated cost
      & \textemdash
      & Run and trajectory component
      & \textemdash
      & Cost composition, growth, and cross-run variation \\
    \midrule
    \multirow{3}{=}{\centering\textbf{Stage 2}\par Utility Attribution}
      & Proxy-Based Attribution
      & IPR~\cite{xiong2024watch}
      & \textemdash
      & Monte Carlo downstream return
      & Intermediate step or action
      & Step-level preference supervision
      & Preference accuracy, return, and task success \\
    \addlinespace[1pt]
      & Information Dependency and Evidence Contribution
      & TRACER~\cite{arxiv2605_09934}
      & \textemdash
      & Claim support and decision-relevant evidence
      & Claim, tool turn, and evidence chain
      & \textemdash
      & Supported, redundant, and unsupported calls \\
    \addlinespace[1pt]
      & Intervention-Based Counterfactual Attribution
      & CAR~\cite{arxiv2606_08275}
      & \textemdash
      & Factual--counterfactual utility difference, $\Delta U(e)$
      & Event or interacting event set
      & \textemdash
      & Intervention effect, confidence interval, and Shapley contribution \\
    \midrule
    \multirow{5}{=}{\centering\textbf{Stages 3--4}\par Diagnosis and Adaptation}
      & Cognitive and Context Misalignment
      & Agent-Omit~\cite{arxiv2602_04284}
      & Input tokens and context occupancy
      & Task result retained after deletion
      & Thought or observation at a turn
      & Dynamic context omission
      & Token reduction and task performance \\
    \addlinespace[1pt]
      & External Interaction Misalignment
      & Budget-Aware Tool-Use~\cite{arxiv2511_17006}
      & Tool-call budget and accumulated path cost
      & Task progress and success under budget
      & Tool call and exploration path
      & Continue, switch path, or stop
      & Success, calls, and budget utilization \\
    \addlinespace[1pt]
      & Recovery-Loop and Verification-Control Misalignment
      & ToolMaze~\cite{arxiv2606_05806}
      & Retry, repair, and verification cost
      & Recovery success after tool faults
      & Fault event and repair loop
      & Fault-aware replanning, bounded retry, and validation
      & Recovery success, retries, and repair cost \\
    \addlinespace[1pt]
      & Resource--Capability Misalignment
      & Ares~\cite{arxiv2603_07915}
      & Reasoning effort and token cost
      & Task reward from allocated effort
      & Individual decision step
      & Assign low or high reasoning effort by step
      & Task success and reasoning cost \\
    \addlinespace[1pt]
      & Multi-Agent Coordination Misalignment
      & AgentPrune~\cite{arxiv2410_02506}
      & Inter-agent messages and model-call cost
      & Task outcome after edge removal
      & Communication edge and agent graph
      & One-shot topology pruning
      & Success, communication volume, and cost reduction \\
    \midrule
    \multirow{4}{=}{\centering\textbf{Stage 5}\par Cost--Utility Evaluation}
      & Cost-Efficiency Metrics and Quality--Cost Trade-offs
      & Terminal-Bench~\cite{arxiv2601_11868}
      & Dollars, tokens, calls, commands, and runtime
      & Executable task and command outcomes
      & Task, trial, turn, and command
      & \textemdash
      & Success, dollars, tokens, calls, and runtime \\
    \addlinespace[1pt]
      & Reliability and Risk-Adjusted Cost Evaluation
      & ReliabilityBench~\cite{arxiv2601_06112}
      & Repeated-run and fault-recovery cost
      & Success under equivalent perturbations and tool faults
      & Configuration, perturbation, and fault level
      & \textemdash
      & Repeated success, robustness, and recovery cost \\
    \addlinespace[1pt]
      & Human and Economic Value
      & GDPval~\cite{arxiv2510_04374}
      & Agent cost, human time, review, and rework
      & Expert-rated occupational deliverable quality
      & Task and end-to-end workflow
      & \textemdash
      & Human time, quality, repair burden, and labor cost \\
    \addlinespace[1pt]
      & Benchmark Infrastructure and Evaluation Protocols
      & BenchGuard~\cite{arxiv2604_24955}
      & \textemdash
      & Validity of benchmark scores and task outcomes
      & Task, reference, environment, and grader
      & \textemdash
      & Consistency defects and evaluation validity \\
    \bottomrule
  \end{tabularx}
\end{table*}

\section{Cost Profiling}
Cost profiling builds the resource side of the alignment framework, identifying which resources LLM agents consume, where consumption occurs, and how costs distribute across tasks, runs, and actions. The output is an auditable ledger that localizes the cost vector of each event, annotates behaviors semantically, and surfaces cost patterns for subsequent utility attribution.

\subsection{Multi-Dimensional Resource Profiling}

Multi-dimensional resource profiling establishes the three fields of the cost vector, covering resource consumption, monetary cost, and risk. In long-horizon agents, context processing, tool invocations, and repair cycles extend the cost boundary beyond a single model inference to a system spanning models, tools, and infrastructure.

Raw token counts undercount agentic cost. In coding agents, historical context re-enters each successive request and the final answer accounts for only a small fraction of total token consumption~\cite{arxiv2604_22750}. Tool invocations add schema injection, transport, and result reinsertion stages that raise token and latency cost beyond call counts~\cite{anjum2026promcp}. Long contexts with serial calls raise KV-cache occupancy and extend latency independently of the token bill~\cite{arxiv2506_04301}.

The monetary field converts resource measurements to deployment-specific cost under fixed model versions, pricing dates, and billing rules, so the same resource usage produces different monetary cost across deployments~\cite{arxiv2511_17006}. The risk field records irreversible externalities such as incorrect writes and data deletion, which consume few tokens but cause disproportionate harm~\cite{ruan2024toolemu}. The three fields are independent, as high resource consumption does not predict monetary cost and low token use does not bound risk.

\subsection{Hierarchical Cost Localization}

Hierarchical cost localization binds resource consumption to the execution structure of an agent, assigning costs to a task, a run, an event, and an executing subject. Task-level totals conflate planning, exploration, tool execution, repair, and verification, while finer granularity identifies where cost concentrates within a run. The executing subject field preserves invocation relationships across agents, enabling attribution in multi-agent systems.

Phase-level decomposition localizes token cost to workflow stages. Tokenomics measures token cost differences across lifecycle phases in software-engineering agents~\cite{arxiv2601_14470}, and Terminal-Bench binds token, latency, and command cost to task, trial, turn, and command records~\cite{arxiv2601_11868}. ClawTrace provides span-level units stable across frameworks, associating token and latency cost with model, tool, and sub-agent calls through subject and parent-child relationships~\cite{arxiv2604_23853}. Across systems, framework-specific labels such as phase, subtask, and step add process semantics within a common task-run-event spine, making each cost vector entry addressable by event and subject.

\subsection{Semantic Action Annotation}

Semantic action annotation separates token cost from risk in the cost vector by classifying actions according to functional role. SWE-agent traces per-step token consumption to interface design, with action granularity reflecting interface choices as much as task content~\cite{arxiv2405_15793}. CostBench finds that the same objective is reachable through different execution sequences, so token counts must be matched against the realized path to measure cost accurately~\cite{liu2026costbench}.

Read and write operations carry identical token counts but differ in risk cost. AppWorld quantifies this asymmetry at the task level, finding that write operations trigger verification, rollback, and downstream token and risk costs absent from read operations~\cite{arxiv2407_18901}. Reversibility determines risk severity, as compensable writes add token overhead for correction while irreversible writes incur irrecoverable risk cost~\cite{arxiv2604_23283}.
\subsection{Workload Characterization and Cost Pattern Analysis}

Workload characterization describes how agent cost distributes across task sets and complete trajectories, spanning composition, temporal evolution, and variation across runs. An agent accumulates cost across retrieval, tool invocation, context expansion, and repair as execution responds to intermediate state.

TraceLab shows that context grows continuously as tool results, inputs, and model outputs accumulate, with cached prefixes accounting for approximately 59.5\% of total estimated cost~\cite{arxiv2606_30560}. Across repeated runs of the same task, model sampling, exploration paths, and repair processes generate occasional high-cost branches, producing substantial token variation~\cite{arxiv2604_22750}.

At the phase level, Tokenomics finds that code review accounts for a major token share through repeated full-code transmission~\cite{arxiv2601_14470}, while Terminal-Bench shows that more tokens or longer trajectories do not yield monotonic success improvements~\cite{arxiv2601_11868}. These patterns locate where cost concentrates or fluctuates, but task contribution must be established from progress signals and counterfactual outcomes.

\section{Utility Attribution}

\begin{figure}[H]
  \centering
  % Editable source: figures/figure3-utility-attribution.drawio.svg
  \includegraphics[width=\columnwidth,keepaspectratio]{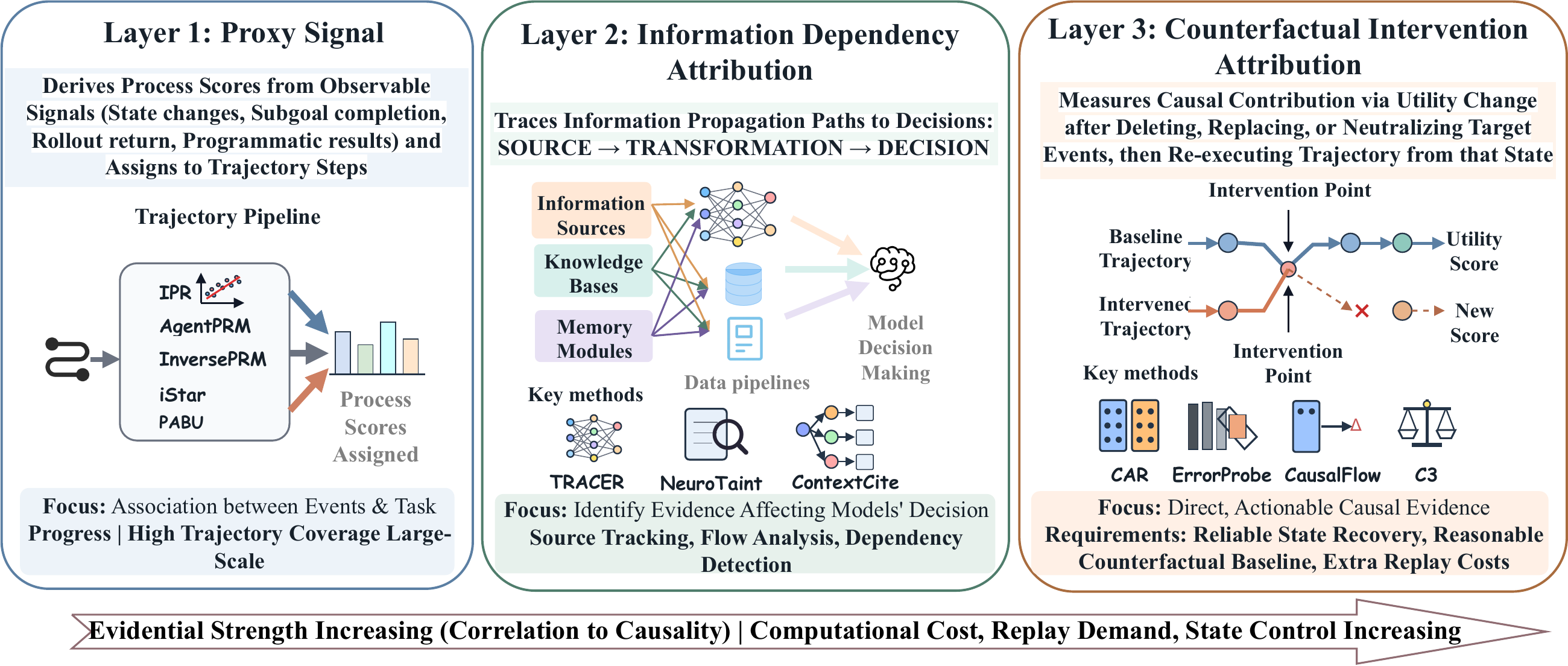}
  \caption{Three-layer utility attribution progressing from proxy signals to information dependencies and counterfactual interventions.}
  \Description{A three-stage view of utility attribution progressing from proxy signals through information dependency to counterfactual intervention.}
  \label{fig:utility-attribution-overview}
\end{figure}

Within the alignment framework, utility attribution constructs the contribution ledger by determining which events, information, and actions in a trajectory causally change the utility signal. Existing methods form an evidence chain of increasing causal strength, from proxy signals through information dependency to counterfactual intervention, as illustrated in Figure~\ref{fig:utility-attribution-overview}.

\subsection{Proxy-Based Attribution}

Proxy-based attribution derives step-level contribution candidates from intermediate trajectory observations including state changes, subgoal completion, programmatic checks, and rollout returns, using proxy utility signals to provide broad trajectory coverage without step-level human annotation.

IPR samples alternative actions at intermediate trajectory points and estimates downstream return as a goal-completion proxy with Monte Carlo rollouts, generating step-level preference pairs that differentiate supervision across failed and successful trajectories~\cite{xiong2024watch}. AgentPRM trains a process reward model to predict each intermediate decision's relationship to long-term return as the utility signal, and its InversePRM variant learns the same utility signal from expert-agent transition comparisons, extending the approach to settings without large-scale rollouts~\cite{arxiv2502_10325}.

iStar and PABU extract step-level signals from existing trajectories without additional environment interaction. iStar recovers action-level goal-completion proxy scores from log-probability ratios between a trained policy and a reference, providing credit assignment without step annotation~\cite{arxiv2509_19199}. PABU estimates step-level partial progress as the utility signal and uses it to compress relevant actions and observations into a compact belief state, jointly supporting local step evaluation and context efficiency~\cite{arxiv2602_09138}. Proxy signals correlate local states with future outcomes but do not establish causal contribution, which requires determining whether the associated evidence actually entered later decisions.

\subsection{Information Dependency and Evidence Contribution}

Information-contribution analysis attributes output quality and constraint satisfaction to specific retrieved content, tool observations, program outputs, and memory, organizing evidence by provenance, information flow, and decision dependency.

TRACER attributes output quality to tool turns by linking answer claims to supporting calls with typed relationships encoding quotation, compression, and inference, separating evidence-bearing calls from redundant ones~\cite{arxiv2605_09934}. NeuroTaint extends attribution to constraint satisfaction, following untrusted sources through reasoning transformation and persistent memory to high-privilege sinks, capturing influence that persists when original wording is rewritten~\cite{arxiv2604_23374}. ContextCite tests context-unit dependency by ablating random subsets and fitting a sparse linear surrogate, with combination ablation revealing redundancy or synergy that single-unit tests miss~\cite{arxiv2409_00729}.

Together they establish information provenance, propagation, and generation dependency, stopping short of whether dependency changes task outcomes.

\subsection{Intervention-Based Counterfactual Attribution}

Intervention-based counterfactual attribution tests the causal role of individual events by deleting, replacing, or neutralizing them, re-executing the trajectory from that point, and computing $\Delta U(e_i)=U(\tau)-U(\tau[e_i{\to}\varnothing])$ between factual and counterfactual conditions. The methods estimate $\Delta U(e_i)$ across single-agent traces, failure-repair scenarios, and cooperative multi-agent systems.

CAR represents actions, observations, and policy as a causal execution process, intervenes on a target node, and re-executes the suffix to measure goal-completion utility, using Shapley-value decomposition to detect redundancy and synergy among events~\cite{arxiv2606_08275}. ErrorProbe localizes failure by tracing dependencies backward from an observable symptom and verifying candidate steps through tool or environment execution against the goal-completion signal~\cite{arxiv2606_01365}. CausalFlow pursues the same goal through step-level direct intervention, replacing a candidate step with a feasible alternative and checking whether the goal-completion outcome flips from failure to success~\cite{arxiv2605_25338}.

C3 extends counterfactual attribution to cooperative multi-agent systems by fixing the interaction history before a target decision, sampling alternative messages or actions for one agent, and evaluating downstream goal-completion effects under matched context~\cite{arxiv2603_06859}. State recoverability bounds comparison precision, as hidden state, irreversible operations, or changing external services require environment snapshots and random seeds beyond what text history can reconstruct. Together they provide the closest approximation to local marginal contribution through executable outcome evidence.

\section{Cost--Utility Misalignment Diagnosis and Targeted Adaptation}

With $C(e_i)$ from Section~3 and $U(\tau)$ from Section~4, this chapter diagnoses where $C(\tau)$ lacks sufficient $U(\tau)$ support and presents targeted adaptations for five forms of misalignment spanning cognitive and context use, external interaction and exploration, recovery-loop and verification control, resource and capability allocation, and multi-agent coordination.

\subsection{Cognitive and Context Misalignment}
\label{subsec:cognitive}

Cognitive and context misalignment arises when token cost and context occupancy grow without changing task-relevant beliefs, action choices, or environmental state. Generation verbosity, unnecessary retention, and redundant retrieval each present a feasible lower-cost alternative that achieves equal utility.

Generation verbosity mismatch occurs when additional tokens produce the same decision. DEPO trains the policy to reach the same decision with less output, reducing token cost per step and total steps to completion, and finds that local verbosity also extends interaction and compounds downstream action cost~\cite{chen2026depo}.

In retention misalignment, completed plans, early observations, and duplicated tool outputs persist beyond their utility contribution. Agent-Omit identifies this by removing a thought or observation and comparing task results and token use, then trains the agent to apply omission dynamically~\cite{arxiv2602_04284}. AgentDiet detects useless, redundant, and expired segments across software-agent trajectories and removes them at inference time, reducing input tokens while preserving task performance~\cite{arxiv2509_23586}.

Naive omission creates a secondary mismatch when it removes content that future steps depend on, reducing current token cost while degrading downstream utility. Beyond Compaction organizes the trace into typed, dependent episodes and evicts by recoverability, prioritizing persisted environment actions and bulk tool output while retaining the objective, active exploration, and referenced nodes, stabilizing the active token count near a fixed ceiling~\cite{arxiv2606_11213}.

PRISM addresses retrieval-reuse mismatch by jointly optimizing retrieval accuracy and context cost through intent-aware graph search and evidence-bundle selection~\cite{arxiv2605_12260}. PEEK resolves context-budget mismatch by maintaining a fixed-budget context map through structured updating and eviction, improving the cost-quality Pareto frontier~\cite{arxiv2605_19932}.

\subsection{External Interaction Misalignment}

External interaction and exploration misalignment centers on whether an agent should access the environment, which tool it should call, and how deeply or broadly it should search. An interaction is misaligned when a no-call, better-tool, or earlier-stop alternative preserves utility without those added costs.

Controlled comparisons reveal a call-task mismatch in which structured knowledge tools reduce error on fact-retrieval but add monetary and latency cost without utility gain on open-ended generation, resolved by gating retrieval and planning to tasks requiring structured evidence or multi-hop composition~\cite{arxiv2601_02663}. Calibrate-Then-Act resolves unnecessary-call mismatch by modeling tool use as a costly partially observable process and gating each call on whether expected information gain would change the chosen action~\cite{arxiv2602_16699}.

INTENT identifies selection mismatch by showing that a cheap but unreliable tool can generate retries and rework exceeding the cost of one expensive stable call, and addresses it through a language world model that anticipates future tool sequences and estimates risk-adjusted cost for each candidate~\cite{liu2026intent}.

Search structure governs how rollout budget distributes across branches and depth beyond the individual call. ToolTree resolves unproductive-branch mismatch by identifying failing sequences through combined pre-execution scoring and post-execution feedback, then applying bidirectional pruning to concentrate budget on calls that survive both screens~\cite{arxiv2603_12740}. W\&D resolves width-depth allocation mismatch, finding that starting with broader parallel calls and narrowing as evidence accumulates outperforms fixed and LLM-controlled scheduling~\cite{arxiv2602_07359}.

Budget exposure and path correction address the accumulated cost of search. Budget-Aware Tool-Use resolves budget-visibility mismatch by showing that increased call limits without budget tracking fail to improve utility, adapting through a Budget Tracker and BATS policy that selects among continuation, path switching, and stopping~\cite{arxiv2511_17006}. WebRollback resolves wrong-path continuation mismatch by letting the agent critique its position and restore an earlier state, stopping cost accumulation and reassigning the remaining budget~\cite{arxiv2504_11788}.

\subsection{Recovery-Loop and Verification-Control Misalignment}

Recovery-loop and verification-control misalignment addresses whether verification, retry, repair, and rollback restore task state after an error, and when the loop should stop. Recovery cost spans verifier calls, error localization, repair generation, and re-verification. Repeated identical failures, patches that leave the failing condition unchanged, and additional rounds without higher success each signal a mismatch, and the objective is to restore correct state at minimum total recovery cost.

ToolMaze and WAREX show through fault injection that a fixed recovery strategy creates a fault-type mismatch, misallocating cost across permanent failures, transient faults, and semantic corruption, each requiring path switching, limited retry, and observation validation respectively~\cite{arxiv2606_05806,kara2025warex}.

Repair spending creates an opportunity-cost mismatch when the same budget applied to independent resampling yields higher utility, making equal-budget resampling the appropriate recovery baseline~\cite{arxiv2306_09896}. Structured feedback including admissible alternatives resolves feedback-quality mismatch, raising repair success substantially over raw error output~\cite{arxiv2607_14167}.

Undetected errors that propagate through the trajectory create an error-propagation mismatch, compounding repair cost while degrading task state. GA-Rollback addresses this by placing an auxiliary reviewer before history accumulation and rolling back to a pre-error state within reversibility bounds~\cite{arxiv2503_02519}. Continued repair past the point of utility gain creates an over-repair mismatch. EET uses historical repair data to terminate patch generation when marginal gain falls below threshold, and VRR-Stop models repair success probability alongside verifier error rates to select among continued repair, acceptance, and stopping~\cite{arxiv2601_05777,arxiv2607_17641}.

\subsection{Resource--Capability Misalignment}

Resource--capability misalignment identifies which model, reasoning intensity, and execution location should perform a step judged worth doing. Excess capacity wastes token and monetary cost on easy steps, while insufficient capability induces rework whose compounded trajectory cost exceeds the per-call savings.

Call-level overprovision mismatch arises when a strong-model inference is assigned to a query where a cheaper model matches utility. Hybrid LLM and RouteLLM identify this through quality-gap prediction and win-probability learning, routing to the cheaper model when the gap is negligible~\cite{arxiv2404_14618,arxiv2406_18665}. BEST-Route resolves sample-count mismatch by identifying when several cheap samples collectively match a single expensive call and when the strong model is required regardless~\cite{arxiv2506_22716}.

Ares addresses step-level effort mismatch by training a policy that assigns low effort to routine navigation and high effort to replanning and recovery, concentrating reasoning cost where it changes outcomes~\cite{arxiv2603_07915}. Inference-Time Distillation resolves effort-concentration mismatch at uncertain states by retrieving teacher traces for a cheaper student and falling back to the teacher when student samples disagree~\cite{arxiv2512_02543}.

Hera identifies placement mismatch by learning step-level device-or-cloud assignments from long-term task reward and future cloud use, keeping simple steps local and routing only pivotal states to the cloud~\cite{arxiv2605_24598}. UniScale resolves joint-configuration mismatch by adapting model size, reasoning effort, sampling, search width, and verification jointly online to task distribution, model availability, and quality targets~\cite{arxiv2605_30898}.

\subsection{Multi-Agent Coordination Misalignment}

Multi-agent coordination misalignment examines whether the arrangement of roles, communication topology, and supervision timing can be reduced without losing utility. Coordination cost spans model calls across agents, shared context, inter-agent messaging, waiting latency, and supervision overhead, accumulating without task progress when roles are redundant, edges carry no decision-relevant information, or supervision fires on already-correct steps.

MAST identifies membership mismatch through failure-trace patterns, and MaAS resolves it by making architecture a task-dependent supernet choice that pays full-team cost only when the task requires it~\cite{arxiv2503_13657,arxiv2502_04180}. Nash-CredMAS resolves contribution-threshold membership mismatch by selecting participants whose expected marginal contribution covers their coordination cost, adapting through context-dependent value estimation under a communication budget~\cite{fan2026nashcredmas}. Shapley-Coop extends membership mismatch resolution to incentive-allocation, linking each member's estimated utility contribution to admission, resource, and cost-sharing decisions under self-interested participation~\cite{arxiv2506_07388}.

AgentPrune identifies topology mismatch when edge density and message similarity reveal removable edges, and prunes them to reduce inter-agent messaging cost without degrading task outcomes~\cite{arxiv2410_02506}. Evolving Orchestration resolves supervision-timing mismatch by invoking each role only during phases where its contribution changes the outcome~\cite{arxiv2505_19591}. SupervisorAgent resolves unnecessary-intervention mismatch through an LLM-free filter that restricts intervention to high-risk or uncertain steps~\cite{lin2026stopwastingtokensefficient}.

For communication judged necessary, repeated encoding of unchanged content across agents and turns wastes compute without utility gain. KVCOMM resolves spatial re-encoding mismatch by aligning and reusing KV cache across agent contexts so that shared content is encoded once~\cite{arxiv2510_12872}. Continuum resolves temporal re-encoding mismatch by managing cache validity across tool waits and session gaps, preventing re-encoding when cached representations remain valid~\cite{arxiv2511_02230}.

\section{Cost--Utility Evaluation}

Cost--utility evaluation closes the alignment loop, testing whether adaptations preserve the utility signal and reduce cost vectors under matched budget constraints. Local compression, routing, or pruning reduces one cost dimension but may raise total expenditure or degrade quality elsewhere. The evaluation takes the agent configuration as its object across complete runs, retaining multi-dimensional cost and utility and reporting raw measures before aggregation to avoid masking cost transfer. The four subsections increase the evidential requirement progressively, with components, questions, and evidence summarized in Table~\ref{tab:cost-utility-evaluation-overview}.

\begin{table}[t]
  \caption{Evaluation components, central questions, and evidence in cost--utility evaluation.}
  \label{tab:cost-utility-evaluation-overview}
  \centering
  \footnotesize
  \setlength{\tabcolsep}{3pt}
  \renewcommand{\arraystretch}{1.15}
  \begin{tabular}{
    >{\centering\arraybackslash}m{0.24\columnwidth}
    >{\centering\arraybackslash}m{0.29\columnwidth}
    >{\centering\arraybackslash}m{0.395\columnwidth}}
    \toprule
    \textbf{Evaluation component} & \textbf{Central question} & \textbf{Key evidence} \\
    \midrule
    Quality--cost trade-offs
      & Is utility preserved per unit cost?
      & Task success, tokens, calls, latency, monetary cost, and Pareto frontier \\
    Reliability and risk
      & Does performance persist under stress?
      & Pass@$k$, Pass$^k$, recovery cost, failure severity, and tail risk \\
    Human and economic value
      & Does deployment reduce total workflow cost?
      & Time saved, output quality, review and rework cost, and net value \\
    Evaluation infrastructure
      & Are comparisons reproducible and auditable?
      & Complete logs, resettable environments, executable scoring, cost logging, and benchmark audits \\
    \bottomrule
  \end{tabular}
\end{table}

\subsection{Cost-Efficiency Metrics and Quality--Cost Trade-offs}

The benchmarks in this subsection examine quality--cost trade-offs under standard conditions, holding task set, model versions, tool permissions, run limits, and prices fixed. Execution cost and action cost are treated as separate dimensions since they differ in unit, location, and bearer, with efficiency ratios summarizing utility per unit resource and budget curves tracing how utility changes across budgets to identify the Pareto frontier.

ScienceAgentBench tracks executable program quality, task success, and API cost, and FDABench extends the same structure to heterogeneous-data analysis with external-model calls and latency~\cite{arxiv2410_05080,arxiv2509_02473}. SWE-Effi and AgencyBench apply the same approach to software engineering and long-horizon tasks, finding that a simpler model--scaffold pairing can match more elaborate configurations at lower cost~\cite{fan2025sweeffi,li2026agencybench}.

Agentless shows that a controlled localization-patching-testing workflow matches more elaborate agents at lower cost~\cite{arxiv2407_01489}. Controlled comparisons find that planning and tool benefits are task-conditional~\cite{arxiv2601_02663}. TPS-Bench finds that improved scheduling simultaneously raises completion rate while reducing tokens, rounds, and time~\cite{arxiv2511_01527}.

CostBench adds environment-action costs and dynamic replanning, finding that action-path expense can substantially exceed the model bill~\cite{liu2026costbench}. CUJBench finds that more tools and evidence can enlarge the search space and lower accuracy, so the quality--cost curve may decline at high budgets~\cite{arxiv2604_23455}. Terminal-Bench integrates all dimensions by jointly logging task outcome, dollars, tokens, calls, and runtime across realistic terminal tasks~\cite{arxiv2601_11868}.

\subsection{Reliability and Risk-Adjusted Cost Evaluation}

Reliability benchmarks evaluate agent behavior under stressed conditions, measuring performance across repeated runs, task-equivalent perturbations, and injected faults. Pass@$k$ and Pass$^k$ distinguish occasional from sustained success across repeat count, perturbation strength, and fault intensity. Risk-adjusted cost incorporates failure severity, recovery expenditure, and catastrophic constraint violation into deployment cost.

$\tau$-bench uses Pass$^k$ to distinguish occasional from sustained success in tool--agent--user interaction~\cite{arxiv2406_12045}. ReliabilityBench extends this to equivalent task perturbations and production-like tool faults, finding that neither larger models nor more complex architectures guarantee greater reliability under stress~\cite{arxiv2601_06112}.

APEX-Agents tracks repeated runs with progress, step, tool-call, and token logging, showing that failed or timed-out trajectories can cost as much as successful ones~\cite{arxiv2601_14242}. WirelessBench adds engineering tolerances and catastrophic-error detection, separating small numerical deviations from critical constraint violations~\cite{arxiv2603_21251}. CCPO minimizes expected deployment cost under a joint reliability threshold and catastrophic-failure bound, allowing configurations to be rejected for instability or tail failure severity even when standard performance targets are met~\cite{si2026ccpo}.

\subsection{Human and Economic Value}

Skilled-human task time, professional output quality, and workflow labor cost form an evidence chain from time saved and expert quality through review and rework burden to risk-adjusted net economic value. A cheaper per-run configuration can increase total workflow expenditure when rework rises, making deployment value dependent on the full chain.

HCAST calibrates the span of agent autonomy against skilled-human completion time across diverse task domains, establishing human baseline time as the measure of task scale and potential value~\cite{arxiv2503_17354}. PaperBench evaluates professional artifact quality through author-informed rubrics for research replication and validates its automatic judge against independent experts~\cite{arxiv2504_01848}.

GDPval combines occupational tasks, expert deliverables, and review and repair workflows, finding that per-run cost and reliability interact to determine total labor cost~\cite{arxiv2510_04374}. SWE-Lancer anchors task value to freelance market payments, finding that full deployment cost requires combining success rates, acceptance criteria, repair cost, and failure risk~\cite{arxiv2502_12115}.

\subsection{Benchmark Infrastructure and Evaluation Protocols}

Credible efficiency, reliability, and economic evaluation depends on the measurement infrastructure supporting it. Observable environments, reproducible initialization, cost-complete run logs, and audited benchmarks and graders form the foundation, spanning fine-grained progress observation, resettable virtual environments, command-level cost traces, and task-level consistency audits.

AgentBoard provides fine-grained progress rates for long-horizon tasks, distinguishing near completion, early failure, and execution without progress when overall success rates are sparse~\cite{arxiv2401_13178}. WebArena, VisualWebArena, and WorkArena extend the environment layer to interactive web and knowledge-work settings with observable state and verifiable task outcomes~\cite{arxiv2307_13854,arxiv2401_13649,arxiv2403_07718}. OSWorld adds virtual-machine initialization and executable state checks, giving configurations and repeated runs a common recoverable starting point for counterfactual replay, perturbation, and fault injection~\cite{arxiv2404_07972}.

SWE-bench and SpreadsheetBench illustrate executable grading for repository patches and spreadsheet outcomes~\cite{arxiv2310_06770,arxiv2406_14991}. Terminal-Bench extends this to cost logging, combining command-level traces, tokens, calls, runtime, and executable tests in a single run record~\cite{arxiv2601_11868}.

BenchGuard audits consistency among task descriptions, reference implementations, execution environments, and scoring logic, finding that apparently inefficient agent runs may reflect defective tasks~\cite{arxiv2604_24955}. PaperBench's JudgeEval compares automatic and human judgments, finding that reliable scoring requires more than a detailed rubric~\cite{arxiv2504_01848}.

\section{Open Challenges}

The five-stage framework reveals four gaps current methods leave unresolved. Attribution requires counterfactual replay that open environments prevent, and no system closes the adaptation loop using attribution evidence while accounting for the cost of obtaining it. Evaluation results do not reproduce across infrastructure snapshots, and existing protocols miss human review overhead and cost transfers outside agent-visible token budgets.

\textbf{Scalable Causal Attribution in Open Environments.} Counterfactual attribution requires replaying execution from a known state, yet open environments prevent this because external writes are irreversible and hidden state cannot be reconstructed from text logs~\cite{arxiv2603_06859}. Multi-agent systems compound the difficulty, as replacing one agent's message changes the information available to all others and single-step deletion methods cannot express joint failures where two steps cause an outcome only together~\cite{arxiv2606_08275}.

\textbf{Closed-Loop Adaptation under Attribution Cost.} Context compression, model routing, tool pruning, and verification control are studied independently~\cite{arxiv2602_09138,arxiv2603_07915,arxiv2410_02506,lin2026stopwastingtokensefficient}, yet in a complete agent they interact, as routing to a weaker model can trigger rework whose compounded cost exceeds the per-call savings~\cite{arxiv2404_14618,arxiv2406_18665}. UniScale jointly adapts configuration controls, but no current system drives adaptation with cost and contribution attribution evidence while enforcing task-level reliability and risk constraints~\cite{arxiv2605_30898}. Diagnostic overhead compounds this problem, since Shapley attribution is exponential in the worst case and a controller must determine whether the expected gain justifies the evidence cost~\cite{arxiv2606_08275,arxiv2605_25338}.

\textbf{Reliability under Nonstationary Infrastructure.} Model snapshots retire, prices and cache rules update, tools modify schemas, and websites update content, preventing benchmark runs from reproducing later~\cite{arxiv2601_06112}. ReliabilityBench finds performance degrades under infrastructure perturbation and BenchGuard finds that apparently inefficient runs may reflect defective tasks~\cite{arxiv2601_06112,arxiv2604_24955}, yet few leaderboards preserve the environment images, model metadata, and replayable logs needed to separate these sources. Production-scale confidence intervals on rare failure modes require 10,000 or more episodes, far exceeding standard benchmark counts and leaving tail-risk claims unsupported~\cite{arxiv2601_06112}.

\textbf{Human Value, Cost Transfer, and Accountability.} Technical evaluations measure resource consumption and task success, but one-shot non-interactive tasks omit human review effort, correction overhead, and failure loss~\cite{arxiv2510_04374,arxiv2503_17354}. GDPval and SWE-Lancer connect agent evaluation to occupational deliverables and market payment rates~\cite{arxiv2510_04374,arxiv2502_12115}, but no study jointly observes agent cost, reliability, and human workflow effort under matched conditions, and existing protocols do not track whether cost optimizations transfer expenses to reviewers.

\section{Conclusion}

We frame LLM agent efficiency as a cost--utility alignment problem: whether an agent's resource consumption is justified by its contribution to task outcomes. We develop a five-stage framework spanning cost profiling, utility attribution, misalignment diagnosis, targeted adaptation, and evaluation, and analyze representative methods at each stage from profiling infrastructure through causal attribution to deployment evaluation. Four open challenges remain: scalable attribution in open environments, closed-loop adaptation under attribution cost, reproducible evaluation under infrastructure drift, and human-centered accounting of labor and cost transfers. Closing them requires methods that connect observable cost records to causal contribution evidence, which current efficiency research does not yet do reliably.

\bibliographystyle{ACM-Reference-Format}
\bibliography{references}

\appendix

\section{TraceCard Schema and Worked Example}

This appendix provides a concrete schema for trajectory-level cost--utility profiling and demonstrates its application through an end-to-end example covering all five framework stages.

\subsection{TraceCard Schema}

Each event $e_i$ in a trajectory $\tau$ carries a \textbf{TraceCard} recording identity, cost profile, attribution evidence, misalignment diagnosis, and adaptation outcome. Table~\ref{tab:tracecard-schema} defines the complete field structure.

\begin{table*}[t]
  \caption{TraceCard schema: field structure per trajectory event.}
  \label{tab:tracecard-schema}
  \centering
  \footnotesize
  \setlength{\tabcolsep}{4pt}
  \renewcommand{\arraystretch}{1.15}
  \begin{tabular}{
    >{\centering\arraybackslash}m{0.12\textwidth}
    >{\centering\arraybackslash}m{0.18\textwidth}
    >{\centering\arraybackslash}m{0.12\textwidth}
    >{\raggedright\arraybackslash}m{0.48\textwidth}}
    \toprule
    \textbf{Field group} & \textbf{Fields} & \textbf{Type / Domain} & \textbf{Purpose and notes} \\
    \midrule
    \textbf{Identity}
      & \texttt{event\_id}
      & string
      & Unique identifier within trajectory \\
    \addlinespace[0.5pt]
      & \texttt{event\_type}
      & enum
      & \{think, call, write, observe, communicate\} — typed action category \\
    \addlinespace[0.5pt]
      & \texttt{actor}
      & string
      & Executing agent or sub-agent identifier \\
    \addlinespace[0.5pt]
      & \texttt{turn}
      & int
      & Sequential position in trajectory \\
    \midrule
    \textbf{Cost Vector}\par\footnotesize (Stage~1)
      & \texttt{tokens\_in}
      & int
      & Input tokens at this event \\
    \addlinespace[0.5pt]
      & \texttt{tokens\_out}
      & int
      & Output tokens generated \\
    \addlinespace[0.5pt]
      & \texttt{latency\_ms}
      & int
      & Wall-clock latency in milliseconds \\
    \addlinespace[0.5pt]
      & \texttt{cost\_usd}
      & float
      & Monetary cost under deployment pricing \\
    \addlinespace[0.5pt]
      & \texttt{risk}
      & enum
      & \{none, low, medium, high, irreversible\} — side-effect severity \\
    \addlinespace[0.5pt]
      & \texttt{reversibility}
      & enum
      & \{reversible, compensable, irreversible\} — recovery feasibility \\
    \midrule
    \textbf{Attribution}\par\footnotesize (Stage~2)
      & \texttt{proxy\_score}
      & float $\in [0,1]$
      & Process-based contribution proxy (§4.1) \\
    \addlinespace[0.5pt]
      & \texttt{dep\_flag}
      & bool
      & Information dependency detected (§4.2) \\
    \addlinespace[0.5pt]
      & \texttt{delta\_U}
      & float | null
      & Counterfactual utility contribution $\Delta U(e_i)$ (§4.3); null when replay unavailable \\
    \addlinespace[0.5pt]
      & \texttt{method}
      & enum
      & \{proxy, dependency, counterfactual\} — evidential strength \\
    \midrule
    \textbf{Misalignment}\par\footnotesize (Stage~3)
      & \texttt{mismatch\_type}
      & enum
      & \{none, cognitive, external, recovery, capability, coordination\} (§5) \\
    \addlinespace[0.5pt]
      & \texttt{severity}
      & enum
      & \{low, medium, high\} — relative cost excess \\
    \addlinespace[0.5pt]
      & \texttt{baseline\_cost\_usd}
      & float
      & Feasible alternative cost for comparison \\
    \midrule
    \textbf{Adaptation}\par\footnotesize (Stage~4)
      & \texttt{adaptation\_method}
      & string | null
      & Applied intervention (e.g., Agent-Omit, DEPO, ToolTree); null if none \\
    \addlinespace[0.5pt]
      & \texttt{cost\_after\_usd}
      & float | null
      & Post-adaptation cost; null before intervention \\
    \bottomrule
  \end{tabular}
\end{table*}

A complete \texttt{TrajectoryRecord} aggregates \texttt{events[]}, \texttt{total\_cost}, task-level \texttt{utility} $U(\tau)$, \texttt{budget} constraint, and \texttt{aligned} boolean indicating whether a lower-cost alternative was feasible.

\subsection{Illustrative TraceCard Instances}

Table~\ref{tab:tracecard-instances} shows two contrasting TraceCard instances from a hypothetical coding-agent trajectory: a high-verbosity reasoning step ($e_{\mathrm{think}}$) and the downstream corrective write step ($e_{\mathrm{write}}$). Values are schematic and chosen to illustrate the contrast between misaligned and aligned events across all five field groups.

\begin{table*}[t]
  \caption{Two contrasting TraceCard instances (illustrative). Bolded cells mark the attribution fields that drive misalignment diagnosis.}
  \label{tab:tracecard-instances}
  \centering
  \footnotesize
  \setlength{\tabcolsep}{5pt}
  \renewcommand{\arraystretch}{1.18}
  \begin{tabular}{
    >{\raggedright\arraybackslash}m{0.28\textwidth}
    >{\centering\arraybackslash}m{0.34\textwidth}
    >{\centering\arraybackslash}m{0.34\textwidth}}
    \toprule
    \textbf{Field} &
    \textbf{$e_{\mathrm{think}}$} &
    \textbf{$e_{\mathrm{write}}$} \\
    \cmidrule(r){2-2}\cmidrule(l){3-3}
    & \textit{verbose reasoning step} & \textit{corrective write step} \\
    \midrule
    \multicolumn{3}{l}{\textit{Identity}} \\
    \quad\texttt{event\_type}   & think        & write \\
    \quad\texttt{actor}         & planner      & executor \\
    \quad\texttt{turn}          & early        & late \\
    \midrule
    \multicolumn{3}{l}{\textit{Cost Vector (Stage~1)}} \\
    \quad\texttt{tokens\_in}    & large (accumulated context) & large (accumulated context) \\
    \quad\texttt{tokens\_out}   & large (verbose generation)  & small (targeted patch) \\
    \quad\texttt{risk}          & none         & medium (file write) \\
    \quad\texttt{reversibility} & reversible   & compensable \\
    \midrule
    \multicolumn{3}{l}{\textit{Attribution (Stage~2)}} \\
    \quad\texttt{proxy\_score}  & low          & high \\
    \quad\texttt{dep\_flag}     & \textbf{false} & \textbf{true} \\
    \quad\texttt{delta\_U}      & $\mathbf{\approx 0}$ (counterfactual replay) & $\mathbf{\approx 1}$ (counterfactual replay) \\
    \quad\texttt{method}        & counterfactual & counterfactual \\
    \midrule
    \multicolumn{3}{l}{\textit{Misalignment (Stage~3)}} \\
    \quad\texttt{mismatch\_type}    & \textbf{cognitive} (§\ref{subsec:cognitive}) & none \\
    \quad\texttt{severity}          & medium      & — \\
    \quad\texttt{baseline\_cost}    & low (direct write feasible) & — \\
    \midrule
    \multicolumn{3}{l}{\textit{Adaptation (Stage~4)}} \\
    \quad\texttt{adaptation\_method} & Agent-Omit~\cite{arxiv2602_04284} & null \\
    \quad\texttt{cost\_after}        & low         & null \\
    \bottomrule
  \end{tabular}
\end{table*}

The contrast illustrates how attribution evidence drives selective diagnosis. $e_{\mathrm{think}}$ carries large output tokens but \texttt{dep\_flag = false} and $\Delta U \approx 0$: counterfactual replay confirms that removing it leaves task outcome unchanged, satisfying the misalignment condition (a feasible lower-cost alternative exists). The diagnosis triggers Agent-Omit~\cite{arxiv2602_04284}, which removes the redundant step while preserving dependency-linked content. $e_{\mathrm{write}}$, by contrast, has \texttt{dep\_flag = true} and $\Delta U \approx 1$: it is causally necessary for task success, so no misalignment is diagnosed and no adaptation is applied. The TraceCard schema makes this asymmetry explicit in a single record, supporting systematic traversal of a full trajectory at Stage~3 and targeted intervention at Stage~4.

\end{document}